\documentclass[aps,prd,onecolumn,nofootinbib,superscriptaddress,showpacs,amsmath,amssymb]{revtex4}
\usepackage{graphicx,epsf}
\usepackage[]{latexsym}
\newcommand{\ve}{\varepsilon}
\newcommand{\be}{\begin{eqnarray}}
\newcommand{\ee}{\end{eqnarray}}
\newcommand{\bea}{\begin{eqnarray}}
\newcommand{\eea}{\end{eqnarray}}

\def\comment#1{}

\newcommand{\lp}{\ell_{P}}
\newcommand{\mpl}{m_{P}}

\newcommand{\om}{\tilde{\omega}}

\usepackage{color}

\definecolor{darkred}{rgb}{.8,0,0}

\definecolor{darkblue}{rgb}{0,0,.7}

\definecolor{darkgreen}{rgb}{0,.7,0}

\begin{document}

%
%
%%%%%%%%%%%%%%%%%%%%%%%%%%%%%%%%%%%%%%%%%%%%%%%%%%%%%%%%%%%%%%
%\title{Generalized uncertainty principle and bounds on Lorentz violation in gravitational sector}
%\title{New bounds on the deformation parameter of GUP from Standard Model Extension}
\title{Planck Stars from a Scale-dependent Gravity theory}

\author{Fabio Scardigli\footnote{corresponding author}}
\email{fabio@phys.ntu.edu.tw}
\affiliation{Dipartimento di Matematica, Politecnico di Milano, Piazza Leonardo da Vinci 32, 20133 Milano, Italy}
\affiliation{Department of Applied Mathematics, University of Waterloo, Ontario N2L 3G1, Canada}
%\affiliation{Yukawa Institute for Theoretical Physics, Kyoto University, Kyoto 606-8502, Japan}
%\affiliation{Department of Applied Mathematics, University of Waterloo, Ontario N2L 3G1, Canada}
%
\author{Gaetano Lambiase}
\email{lambiase@sa.infn.it}
\affiliation{Dipartimento di Fisica "E.R. Caianiello", Universita' di Salerno, I-84084 Fisciano (Sa), Italy \&\\
INFN - Gruppo Collegato di Salerno, Italy}
%

%\affiliation{}
%\date{\today}
\begin{abstract}
\par\noindent
Scale dependence of fundamental physical parameters is a generic feature of ordinary quantum field theory. When applied to gravity, this idea produces effective actions generically containing a running Newtonian coupling constant, from which new (spherically symmetric) black hole spacetimes can be inferred.  
%The Asymptotically safe gravity program suggests a specific form of the running Newtonian coupling constant which depends on two free parameters, usually denoted with $\om$ and $\gamma$. New metrics can be inferred from a "running" gravitational constant,. 
As a minimum useful requirement, of course, the new metrics should match with a Schwarzschild field at large radial coordinate. By further imposing to the new scale dependent metric the simple request of matching with the Donoghue quantum corrected potential, we find a not yet explored black hole spacetime, which naturally turns out to describe the so-called Planck stars.    

%\textbf{\color{blue}
%ABSTRACT DA SCRIVERE - Safety Gravity - 
%We present a procedure to link the deformation parameter $\beta$ of the generalized uncertainty principle (GUP) to the two free parameters $\om$ and $\gamma$ of the running Newtonian coupling constant of the Asymptotic Safety gravity (ASG) program. To this aim, we compute the Hawking temperature of a black hole in two different ways. The first way involves the use of the GUP in place of the Heisenberg uncertainty relations, and therefore we get a deformed Hawking temperature containing the parameter $\beta$. The second way involves the deformation of the Schwarzschild metric due to the Newtonian coupling constant running according to the AS gravity prescription. The comparison of the two techniques yields a relation between $\beta$ and $\om$, $\gamma$. 
%}
\end{abstract}
%
%\pacs{03.65.Ta, 03.65-w}

%
\maketitle
%%%%%%%%%%%%%%%
\section{Introduction}
%\label{intro}

%\textbf{\color{blue} RIPARAFRASARE (CUT \& PASTE DA ARTICLO PER FISSARE REFERENCES E IDEA GENERALE) - 
In many of the existing approaches to quantum gravity (for an incomplete list see e.g. \cite{QG1,QG2,QG3,QG4,QG6,QG8,QG9} and references therein) the fundamental parameters that enter the action, such as Newton's constant, electromagnetic coupling, the cosmological constant etc., become scale dependent quantities. This looks quite natural, since scale dependence at the level of the effective action is a generic feature of ordinary quantum field theory. In particular, in theories of gravity, the scale dependence is expected to modify the horizon, the thermodynamics, the quasinormal modes spectra of classical black hole backgrounds 
\cite{Koch:2016uso,
%Rincon:2017ypd,
Rincon:2017goj,
%Rincon:2017ayr,
Contreras:2017eza,Rincon:2018sgd,Contreras:2018dhs,
%Rincon:2018lyd,
%Rincon:2018dsq,
%Contreras:2018gct,
Rincon:2019cix,
%Contreras:2019fwu,
%Contreras:2019cmf,
Contreras:2018gpl}. 
%Also, the Sagnac effect \cite{Rincon:2019zxk}, the evolution of trajectories of photons \cite{Fathi:2019jid}, some cosmological solutions \cite{Canales:2018tbn}, and transverse wormhole solutions \cite{Contreras:2018swc} have been investigated.
%
%In the group of theories beyond classical GR, among the aforementioned 
In the large class of approaches based on scale-dependent gravity (SDG), among the theories beyond classical general relativity (GR), we find also the particular method usually known as asymptotically safe gravity (ASG) \cite{Bonanno:2000ep,Bonanno:2001xi,Reuter:2003ca,
%}. For the renormalization group (RG)-improved black hole metrics, cosmologies, and inflationary models from asymptotic safety, the reader can usefully consult e.g. Refs.~\cite{
Platania:2020lqb,Bonanno:2002zb,Bonanno:2001hi,Liu:2018hno,Hindmarsh:2012rc,Platania:2019kyx,Moti:2018uvl,Koch:2013owa,Koch:2014cqa,Saueressig:2015xua,Bonanno:2016dyv,chinosII,Li:2013kkb}.
%Pawlowski:2018swz,Gonzalez:2015upa}.
%
%\smallskip
%
In this scenario the Newton's constant $G$ is actually seen as a "running" Newton's coupling $G(k)$, which is constructed by integrating the differential equation for the beta function of the gravitational
coupling. Therefore $G$ depends on some arbitrary renormalization energy scale $k$. Furthermore, in concrete black hole models, usually a link between the energy scale $k$ and the radial coordinate $r$ is established. Finally, the composite function $G(k(r))$ is inserted back into the classical black hole solution. A modified lapse function is thus obtained, and these modified black hole metrics are thought to automatically incorporate and describe the effects of quantum gravity. 
Other different black hole metrics, in particular metrics not affected by a central singularity, can be found e.g. in Refs.~\cite{bardeen,hayward,flachi,frolov,BS2}, although those examples may not be directly connected with a SDG program.\\
Generic examples of SDG theories, in particular scale-dependent black hole solutions, can be found in literature, see for example Refs.~\cite{Koch:2016uso,Contreras:2017eza,Contreras:2018gpl}. As for the subclass ASG, a specific well known example (see e.g. Refs.~\cite{Bonanno:2000ep,LS}) is defined by the "running" Newton's coupling $G(k)$ and function $k(r)$ 
\be
\label{Gk}
G(k)= \frac{G}{1+ \om G k^2/\hbar} \ \quad \quad {\rm with} \quad \quad k(r) = \hbar\left(\frac{r+\gamma GM}{r^3} \right)^{1/2}
\ee
where we take the conventions $c=k_B=1$ and we retain $\hbar$. Here $\om$ and $\gamma$ are dimensionless numerical parameters. Combined together the above functions yield the scale-dependent Newton's constant
\be
G(r) = \frac{G r^3}{r^3 + \om G \hbar \left(r + \gamma G M \right)} \,.
\label{Geffr}
\ee 
In the context of ASG literature, the parameter $\om$ is always assumed to be \textit{positive}, $\om >0$. 
In fact, considering the running Newton's coupling $G(k)$ in Eq.\eqref{Gk}, we see that $dG/dk<0$ only if $\om>0$. In other words, in order to keep gravity "asymptotically safe" from divergences, we need to have not only 
$G(k) \to 0$ for $k \to \infty$, but also $G(k) \geq 0$ and \textit{finite} for any $k \geq 0$. Therefore, only when $\om>0$ we are properly in the realm of an Asymptotically Safe Gravity theory. 

On the contrary, if we allow $\om<0$, then $G(k)$ diverges for 
$k \sim k_{Planck}$, becomes negative for $k > k_{Planck}$ and $G(k) \to 0^-$ when $k \to \infty$. This behavior of 
$G(k)$ is clearly outside the specific spirit of ASG theory. Nevertheless, a theory with $\om <0$ can still be classified as a scale-dependent gravity (SDG) theory. In the following we will see that this is actually our present case. \\

A different, well known research line, pursued by John Donoghue and other authors~\cite{Duff,donoghue,hamber,kiril,dono2,kiril2,dono3,donoASP,frob}, considers the reformulation of General Relativity as an effective quantum field theory of gravity at low energies. Along the years, these authors established a solid prediction of the quantum corrections to the Schwarzschild metric, and therefore to the Newtonian potential, at least at the first order in $\hbar$. 
According to a vast literature, a way to determine the free parameters appearing in a SDG theory, and in particular in a ASG-metric (see e.g. Refs.~\cite{Bonanno:2000ep,Bonanno:2001xi,Reuter:2003ca,Platania:2020lqb,Bonanno:2002zb,Bonanno:2001hi,Liu:2018hno,Hindmarsh:2012rc,Platania:2019kyx,Moti:2018uvl,Koch:2013owa,Koch:2014cqa,Saueressig:2015xua,Bonanno:2016dyv,chinosII,Li:2013kkb}), consists in comparing the effective Newtonian potential predicted by the SDG/ASG approach with the one computed in the framework of GR as an effective QFT. 
 By following that path, we are led to investigate, for the first time, a \textit{negative} value for the parameter 
$\om$, unlike previous early studies (see e.g. Refs.~\cite{Bonanno:2000ep,Koch:2014cqa})
\footnote{Here we should at least mention the ongoing lively debate among the communities working on GR as an Effective Field Theory (EFT), and those on the ASG approach. The focus is about "if" and "to what extent" the results obtained on ASG-improved metrics by the ASG community can be compared with the results on the quantum corrected metrics obtained by the GR-EFT community (see e.g. Refs.~\cite{donoASP,BonannoFP}). Here, of course, we cannot enter in details. However, after a long period during which this matching has been carried out almost routinely, without particular problems, 
%However, within the Asymptotic Safety literature the parameter 
%$\om$ is usually considered positive. In fact, considering (see e.g. Refs.~\cite{Bonanno:2000ep,LS}) the running Newton's coupling $G(k)=G/(1+ \om G k^2/\hbar)$, where $k$ is the energy or momentum scale, we see that $dG/dk<0$ only if $\om>0$. With an $\om<0$, $G(k)$ diverges for $k \sim k_{Planck}$, and this behavior of $G(k)$ looks, in principle, outside the spirit of ASG.\\
%On the other hand, while the calculations of the EFT community point nowadays firmly towards a positive coefficient of $\hbar$, 
quite recently, doubts have been risen about the effective comparability of the calculations of the two communities, and the conceptual correctness of such comparison. 
%and therefore on the possibility to infer from that comparison a negative value for $\om$. 
In particular, it would seem that the two different frameworks effectively consider, and implement, different classes of Feynman diagrams in order to get their final (quantum corrected) metric expressions. Of course, including or not including certain subclasses of diagrams could change the sign of the parameters involved. Thus, the overall feeling is that at present a comparison between GR-EFT and ASG results could even be not well grounded, and in any case the available calculations are not yet able to clarify the situation.
This however looks to us as an even stronger motivation to explore the physical consequences of a \textit{negative} 
$\om$ parameter.}. 

As said above, this classifies our approach outside a standard ASG theory, instead closer to a more general SDG theory. Quite surprisingly, we will see that the negative $\om$ leads directly, without further assumptions, to a specific metric which is able to describe important features of the so called Planck stars. It is remarkable that, while Planck stars were introduced in Ref.~\cite{RovelliPS} on the grounds of plausible, general physical considerations, here on the contrary they appear as a possible direct consequence of a Scale-Dependent Gravity approach. \\        

%The aim of this paper is to present and discuss a link between the free parameters of ASG, i.e. the renormalization scale, and the deforming parameter $\beta$ of the generalized uncertainty principle (GUP) . 
%%We shall consider the deformation of Heisenberg uncertainty principle (HUP), or Generalized Uncertainty Principle GUP, of the form
%%
%%$\Delta x \Delta p \gtrsim\frac{\hbar}{2}(1+\beta \frac{4 l_p^2}{\hbar^2}\delta p^2)$,
%%in particular to the deforming parameter of the GUP, i.e. $\beta$ \cite{GUPearly}.
%%
%Here we shall use the GUP to compute the Hawking temperature of a given black hole, obtaining it as a function of the deformation parameter $\beta$. The same deformed temperature can however be computed as well via the ASG deformation of Schwarzschild metric. This will yield a relation between $\beta$ and the ASG free parameters.

The paper is organized as follows. In the next Section we shortly recall the main results leading to a SDG/ASG-modification of the Schwarzschild metric. 
%In Section 3 we compute the GUP-deformed Hawking temperature. In Section 4 we compute the ASG-deformed Hawking temperature. In Section 5 we compare such two deformations and relate the respective deformation parameters. 
In Section III we give a precise evaluation of the SDG/ASG parameters. Section IV is devoted to discuss the features of the SDG-modified Schwarzschild metric, and then Section V discusses the "prediction" of the Planck star model. In Sections VI, VII, VIII, IX we study, respectively, temperature, specific heat, emission rate equation, and thermodynamic entropy of this SDG-modified Schwarzschild/Planck star metric. The last Section contains some quick hints to the phenomenology of these objects (and related references), and the conclusions.

Here we work in units where $c=k_B=1$, and the Planck length $\lp$ is defined as $G\hbar=\lp^2$. Then of course the Planck mass $\mpl$ satisfies $2G \mpl=\lp$ and $\hbar=2\mpl\lp$.

%%%%%%%%%%%%%%%%%%%%%%%%%%%%%%%%%%%%%%%%%%%%%%%%%%%%%%%%%%%%%%%%%%%%%%%%%
\section{Black hole metrics from Running Newtonian coupling} 
\label{Classical}
\noindent
%%%%%%%%%%%%%%%%%%%%%%%%%%%%%%%%%%%%%%%%%%%%%%%%%%%%%%%%%%%%%%%%%%%%%%%%

As we said, according with the literature on the topic, once constructed the running coupling $G(r)$ (through the steps detailed e.g. in Refs.~\cite{Bonanno:2000ep,LS}) 
%in particular Ref.~\cite{Bonanno:2000ep}, we can say that the main steps towards the construction of a renormalization group improved Schwarzschild metric are essentially three: first, we integrate the beta function for the gravitational coupling to compute Newton's constant as a function of some energy scale $k$, namely $G(k)$. After that, a link between the energy scale $k$ and the radial coordinate $r$ must be established, the so called "identification of the infrared cutoff", namely $k=k(r)$. Finally, the $G(r)$ Newton's constant so obtained is inserted into the classical black hole solution and we arrive to an improved lapse function of the metric. Only after this final step the complete metric of a "renormalization improved" black hole becomes concretely usable for explicit calculations. The above steps are detailed, for example, in Refs.~\cite{Bonanno:2000ep,LS}.
the basic idea of the SD/AS gravity approach in order to obtain a 'renormalization improved' (classical Newtonian, or general relativistic) solution is to replace everywhere the numerical Newton constant $G$ with the 'running constant' $G(r)$, whose explicit form is given by Eq.\eqref{Geffr}.
%\cite{Bonanno:2000ep}
%\be
%\label{Geffr}
%G(r) = \frac{G r^3}{r^3 + \om G \hbar \left(r + \gamma G M \right)} \,,
%\ee
%where, in accordance with our conventions, $c=k_B=1$ and we retained $\hbar$. Here $\om$ and $\gamma$ are dimensionless numerical parameters, whose concrete value will be discussed later.
 
The line element for the spherically symmetric, Lorentzian metric preserves the usual form, that is 
\be
\mathrm{d}s^2 = F(r)\mathrm{d} t^2 - F(r)^{-1} \mathrm{d} r^2 - r^2 \: \mathrm{d}\Omega^2,
\label{metric}
\ee
where $r$ is the radial coordinate, and $\mathrm{d} \Omega^2 = \mathrm{d \theta^2 + sin^2 \theta \: d \phi^2}$ is the line element of the unit two-sphere. But now, according to the above prescriptions, the lapse function $F(r)$ of our SDG-improved Schwarzschild geometry reads
\be
\label{ASGmetric}
F(r) = 1 - \frac{2 M G(r)}{r}=1-\frac{2 G M r^2}{r^3 +  \om G \hbar (r + \gamma G M)}\,,
\ee
with $G(r)$ given by (\ref{Geffr}) and $M$ the mass of the black hole. Of course, we suppose $\om \neq 0$, otherwise we would go back to the standard Schwarzschild metric. 
Two very important limiting cases should be considered. \\
The first corresponds to the low energy scales ($r \rightarrow \infty$, or $k \to 0$), which implies 
\be
F(r \rightarrow \infty) \simeq 1 - \frac{2 G M}{r} , 
\label{infty}
\ee
so the standard Schwarzschild metric at large distances is recovered, and this behavior is independent from the values of $\om$ and $\gamma$. \\ 
The second corresponds to the high energy scales ($r \rightarrow 0$, or $k \to \infty$). Here we have to distinguish two subcases.\\
If $\gamma \ne 0$, then  
\be
F(r \rightarrow 0) \simeq 1 - \frac{2 r^2}{\om \gamma G \hbar}\,,
\label{zero}
\ee
and thus the lapse function corresponds to a deSitter ($\om\gamma > 0$) or an Anti-deSitter ($\om\gamma <0$) core of our metric, depending on the sign of $\om\gamma$.\\ 
If $\gamma = 0$, then 
\be
%\label{ASGmetric}
F(r) = 1-\frac{2 G M r}{r^2 +  \om G \hbar}\,,
\ee
and therefore
\be
F(r \rightarrow 0) \simeq 1 - \frac{2 M r}{\om \hbar}\,,
\ee
so in this case we have a conic singularity at the origin.
Clearly, the presence of $\hbar$ signals the quantum character of the correction that the SDG/ASG approach gives to the core of the standard Schwarzschild metric. In both cases the central singularity has disappeared.

\section{Possible values of the parameters $\om$ and $\gamma$}
%%%%%%%%%%%%%%%%%%%%%%%%%%%%%%%%%%%%%

As we said in Sec.II, the SDG/ASG-modified Newtonian potential can be obtained from the standard Newton formula
\be
V(r) = - \frac{G M m}{r}
\ee
by simply replacing the experimentally observed Newton constant $G$ with the running coupling $G(r)$ given in Eq.\eqref{Geffr}. Thus we get
\be
V^{SDG}(r) = -\frac{G(r)Mm}{r} = -\frac{G \,M\, m\, r^2}{r^3 + \om \,G \,\hbar \,(r + \gamma G M)}\,,
\ee
which can be expanded for large $r$ as
\be
\label{Newrun}
V^{SDG}(r) = - \frac{G M m}{r}\left[1 - \frac{\om G \hbar}{r^2} - \frac{\gamma \om G^2 \hbar M}{r^3} + {\cal O}\left(\frac{G^2 \hbar^2}{r^4}\right)\right]\,.
\ee
Clearly, the corrections to the standard Newtonian potential predicted by the SDG/ASG approach are all of quantum nature. This is suggested by the presence of $\hbar$ in each term of correction. In fact, there are no correction terms of classical origin, coming from some kind of post-Newtonian approximation.    
On the other hand, corrections of quantum origin to the classical Newtonian potential have been elaborated by several researchers \cite{Duff,donoghue,hamber,kiril,dono2,kiril2,dono3,donoASP,frob} in the last three decades or so. In particular, it was pointed out by Donoghue \cite{donoghue,dono2} that the standard perturbative quantization of Einstein gravity leads to a well defined, finite prediction for the leading large distance quantum correction to Newtonian potential (the Effective Field Theory approach mentioned in the Introduction). The numerical coefficients of the quantum expansion have undergone a certain evolution over the years~\cite{hamber,kiril}, but the result today accepted by the community~\cite{dono2,kiril2} reads  
\be
\label{Newquan}
  V^{QGR}(r) \ = \
	- \frac{G M m}{r}\left[1 \ + \ \frac{41}{10\pi}\frac{G \hbar}{r^2} \ + \ \dots\right]\,.
\ee
This is an expansion at first order in $\hbar$ (and at second order in $1/r$), 
%where the first correction term does not contain any power of $\hbar$, namely it is a purely classical effect due to non-linear nature of General Relativity, while 
where the first correction term represents a genuine quantum correction proportional to $\hbar$.

The comparison of the two expansions \eqref{Newrun} and \eqref{Newquan} allows us to fix the parameter $\om$, which results to be
\be
\om = - \frac{41}{10\pi}\,.
\label{om}
\ee 
The parameter $\gamma$ cannot be fixed by these considerations. To this aim, we refer the reader to the arguments originally developed in Ref.\cite{Bonanno:2000ep}, and then taken up also by other authors (e.g.\cite{Koch:2014cqa}). Those classical general relativistic arguments 
%have to do with the correct identification of the infrared cutoff, and they 
fix $\gamma = 9/2$. Different kind of considerations, based on the generalized uncertainty principle (see Ref.~\cite{LS}; see also Refs.~\cite{GUP}), lead to the value $\gamma=0$. In this paper we will assume always $\gamma \geq 0$, and in some specific cases we shall comment on the special value $\gamma=0$. However, most of the results will be qualitatively the same for all $\gamma > 0$. \\  
%Using this value of $\gamma$ and the Eqs.\eqref{beta},\eqref{om}, we can compute the $\beta$ parameter of the GUP, which results to be
%\be
%\beta = \frac{451 \, \pi}{5}\,.
%\ee
%As we see $\beta$ results to be of order $10^2$ as suggested by some models of string theory \cite{VenezGrossMende}.
%\item[$ii)$] We can first compute $\beta$, by following the procedure described in Ref.\cite{SLV}, namely comparing the GUP Hawking temperature \eqref{Tg} with the Hawking temperature of a Schwarzschild metric corrected through the Donoghue potential \eqref{Newquan}. This yields $\beta=82\pi/5$. Then, using \eqref{beta},\eqref{om}, we get
%\be
%\gamma = 0\,.
%\ee 
%\item[$iii)$] Finally, we can also adopt the values $\beta=82\pi/5$, and $\gamma = 9/2$, which, via Eq.\eqref{beta}, bring to the value
%\be
%\om = -\frac{41}{55\pi}\,.
%\ee
%\end{description}

%The study of the metric \eqref{ASGmetric} and of the related black hole features (in particular, for $\om<0$) will be carried out in a companion paper. 

As we said in footnote 1, the last couple of years have witnessed the raising of some doubts against the conceptual correctness of a comparison between the SDG/ASG-modified Newtonian potential \eqref{Newrun}, with the GR-EFT quantum corrected Newtonian potential \eqref{Newquan}. Without supporting here one side or the other of this not yet settled down question, it is however perfectly legitimate to explore the physical consequences of the case $\om < 0$, as on the other hand the above Eq.\eqref{om} strongly suggests.  

We emphasize that the \textit{sign} of $\om$ is crucial for the physics of SDG-modified black hole. In the standard, strict, ASG approach, only metrics with $\om>0$ have been studied. This choice of course has implied black holes where the central singularity is wiped out, in favour of a De Sitter or an Anti De Sitter core, as can be easily inferred from Eqs.\eqref{ASGmetric}\eqref{zero}, and is also widely discussed in the above References. In the next sections, on the contrary, starting from a more general SDG point of view, we shall explore metrics with a \textit{negative} $\om$. This choice, as we will see, will have deep consequences on the structure and the physics of the black hole metric \eqref{ASGmetric} considered  
\footnote{From a historical point of view, it is interesting to note that, until quite recently, many authors of ASG community (e.g.\cite{Bonanno:2000ep,Koch:2014cqa,Saueressig:2015xua}), in order to fix the $\om$ parameter, have relied on a direct comparison  between the two expressions \eqref{Newrun} and \eqref{Newquan}. The sign of the $\om$ parameter has therefore changed according to the sign of the first order correction term in $\hbar$ in the expansion \eqref{Newquan}. For example, Duff, in his first calculation \cite{Duff} of 1974 obtained an expansion of the same kind of \eqref{Newquan}, with a \textit{positive} coefficient of the $\hbar$ term (and therefore a negative $\om$). The ASG community has instead actually considered, until few yars ago, the early calculations performed by Donoghue and others~\cite{donoghue,hamber} in the period 1994-1995, where the $\hbar$ coefficient obtained was negative. As a consequence, the ASG authors got a \textit{positive} value of $\om$. However, during the years, the analytical techniques used in GR as an effective QFT have been refined, and the results now accepted by the GR-EFT community are those expressed, initially, in Refs.\cite{dono2,kiril2}, and then confirmed in Refs.\cite{dono3}, as well as in the  very recent Ref.\cite{frob}. All these results coherently point to a \textit{positive} $\hbar$ coefficient in the expansion \eqref{Newquan} (and hence to a \textit{negative} value of $\om$, if the above comparison is effectively considered).}.

\section{Study of the new SDG-modified Schwarzschild metric}
%\section*{Planck stars from Asymptotic safety gravity?}
%%%%%%%%%%%%%%%%%%%%%%%%

\begin{figure}[t]
	%\hspace{-2mm}
	\includegraphics[scale=0.4]{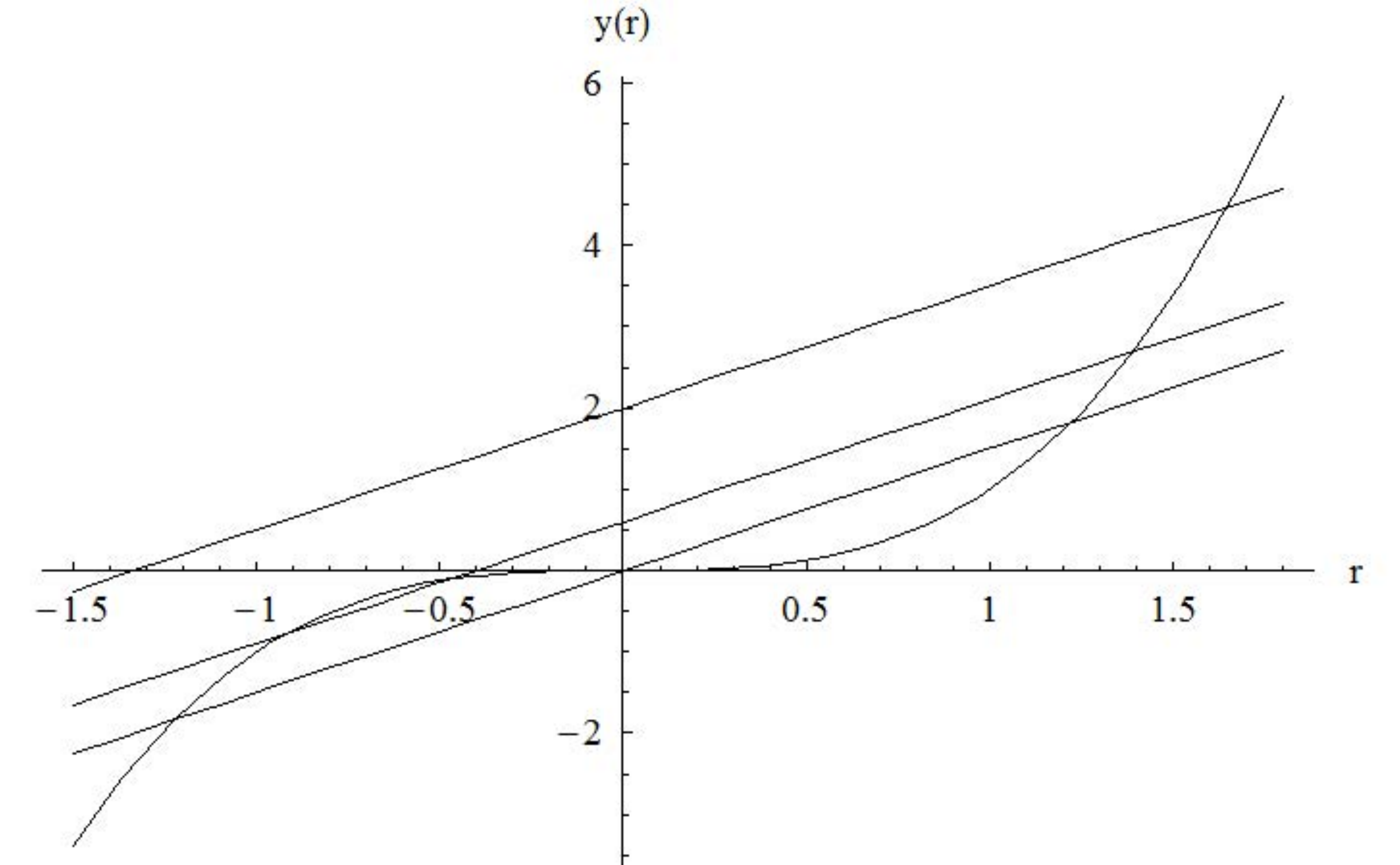}
	\centering
	\protect\caption{Comparison of the two lines $y_1(r)$, $y_2(r,M)$ for increasing values of $M \geq 0$. See Eqs.\eqref{Y}.}
	%$2\pi E/\hbar=1$ and $\beta \ell_p^2=1$ for simplicity.}
	\label{fig1}
\end{figure}

\begin{figure}[b]
	%\hspace{-2mm}
	\includegraphics[scale=.4]{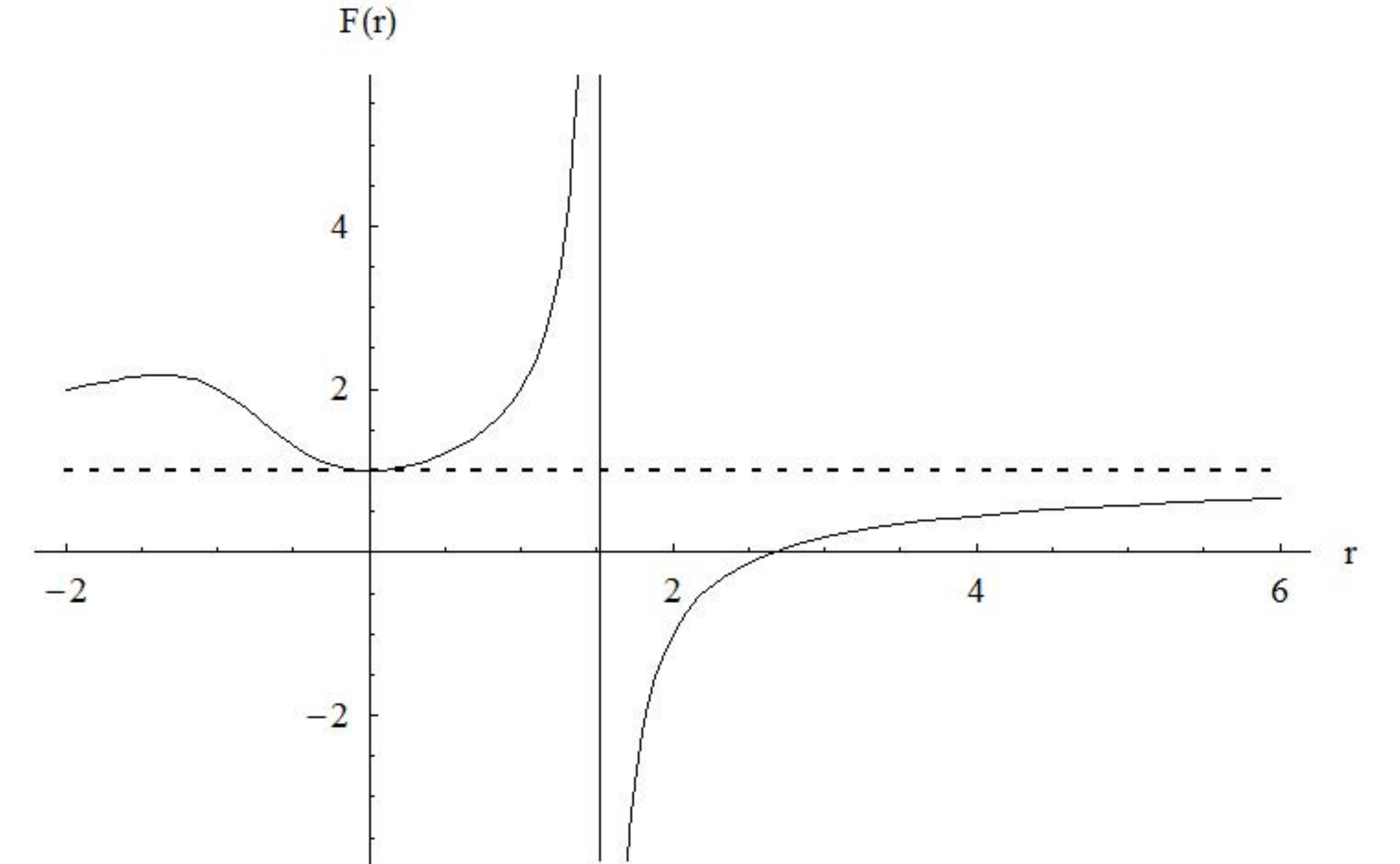}
	\centering
	\protect\caption{Lapse function $F(r)$: physical region for $r>0$, singularity at $r=r_0$, horizon at $r=r_h$, where $F(r_h)=0$.}
	%$2\pi E/\hbar=1$ and $\beta \ell_p^2=1$ for simplicity.}
	\label{fig2}
\end{figure}

The key suggestion coming from the previous Section is to consider the possibility of a \textit{negative} $\om$. This, as we will see, represents a major change in respect to others modified (but regular) Schwarzschild metrics present in literature~\cite{Contreras:2017eza,Contreras:2018gpl,Bonanno:2000ep,hayward,flachi,frolov}. Instead, some contact with our results can be found in Ref.~\cite{Bargueno:2016qhu}, although there the authors don't deal with SDG models. So, according to the previous section, we consider here the case
\be
\om < 0 \ \Rightarrow \ \om = -|\om|\,; \quad \quad \quad \gamma>0\,.
\ee 
The lapse function \eqref{ASGmetric} can therefore be written as 
\be
F(r) = 1-\frac{2 G M r^2}{r^3 -  |\om| G \hbar (r + \gamma G M)}\,.
\label{negom}
\ee
While for $\om>0$ the lapse \eqref{ASGmetric} is regular everywhere when $r>0$ (see e.g. Ref.~\cite{Bonanno:2000ep}), here, with $\om<0$, the scenario is very different. First, we notice that the behavior of $F(r)$ at $r\to \infty$ remains that described by Eq.\eqref{infty}, namely a standard Schwarzschild metric for large $r$. At $r\to 0$ we have an Anti-DeSitter core, namely $F(r) \simeq 1 + 2r^2/(|\om|\gamma G \hbar)$. But now the denominator $D(r)$ appearing in \eqref{negom} can develop zeros. Luckly, a simple graphical analysis is sufficient to clarify the situation. In Fig.\ref{fig1}, we compare the two lines 
\be
y_1 & = & r^3 \nonumber \\
y_2 & = & |\om| G \hbar (r + \gamma G M)
\label{Y}
\ee
%%%%%%%%%%%%%%%%%%%%%%%%%%%%%
%
%%%%%%%%%%%%%%%%%%%%%%%%%%%%%%%
for various values of $M>0$, the mass of the central body. As we see, since $D=y_1-y_2$, there is always only one single zero for $D(r)$ when $r>0$, let's call it $r_0$. So
\be
D(r) & > & 0 \quad {\rm for} \quad r>r_0 \nonumber \\
D(r) & = & 0 \quad {\rm for} \quad r=r_0 \nonumber \\
D(r) & < & 0 \quad {\rm for} \quad 0<r<r_0 \nonumber 
\ee 
In the unphysical region $r<0$, $D(r)$ can develop two distinct zeros, or two coincident zeros, or no real zeros at all. On the ground of the above situation, we can develop a straightforward analysis of the function $F(r)$, valid for any $M>0$:
\be
&&\lim_{r\to +\infty}F(r) = 1^-\,; \quad \lim_{r\to r_0^+} F(r) = 1 - \frac{2G M r_0^2}{0^+} = -\infty\,;\nonumber \\
&&\lim_{r\to r_0^-} F(r) = 1 - \frac{2G M r_0^2}{0^-} = +\infty\,; \quad \lim_{r\to 0^+}F(r) = 1^+\,;\nonumber \\
&&F(r) \sim 1 + \frac{2r^2}{|\om|\gamma G \hbar} \quad {\rm for} \quad r\sim 0\,;
\quad F'(r)=\frac{2G M(r^4+|\om|G \hbar r^2+2|\om|\gamma G^2 M\hbar r)}{(r^3 -  |\om| G \hbar r - |\om|\gamma G^2 \hbar M)^2} > 0  \quad {\rm for} \quad r> 0    \nonumber
\ee  
The information collected above allows us to draw a general graph (Fig.\ref{fig2}) of the lapse function $F(r)$ in the region $r>0$, valid for any $\gamma >0$ and of course $M>0$.
%%%%%%%%%%%%%%%%%%%%%%%%%%%%%
%
%%%%%%%%%%%%%%%%%%%%%%%%%%%%%%%
As we see, there is one single positive zero $r=r_h$ of $F(r)$, which is the horizon of the SDG-modified black hole metric \eqref{negom}. There is also an essential singularity at $r=r_0>0$. The most relevant difference with the standard Schwarzschild metric is the fact that the essential (ineliminable) singularity is at $r_0>0$, instead of being at $r_0=0$. It is also clear that it is always $r_0<r_h$, for any $M>0$. So the singularity is always protected by the event horizon. The singularity is never naked, in full accordance with the Cosmic Censorship Conjecture. 

We now examine the asymptotic behavior of the horizon $r_h$ and of the singularity $r_0$ in the physical relevant limits of large $M$ and small $M$.

%%%%%%%%%%%%%%%%%%%%%%
\subsection{Horizon $r_h$}
%%%%%%%%%%%%%%%%%%%%%%

The only real positive zero of $F(r)$, the horizon, must be a solution $r=r_h$ of the equation
\be
r^3 -2G M r^2 - |\om|G\hbar\, r - \gamma |\om|G^2\hbar M = 0\,.
\label{H}
\ee
The classical limit, $\hbar \to 0$, of the above equation reads: $r^3 -2G M r^2 = 0$,
whose positive solution is $r=2G M$. Therefore, to get an approximate solution of \eqref{H} for large $M$, we pose
\be
r=2G M + \ve 
\ee 
and we perturb around the classical solution $2G M$ by keeping only the first order terms in $\ve$. 
\footnote{An alternative derivation is the following. Since $G\hbar = \lp^2$, equation \eqref{H} reads: $r^3 -2G M r^2 - |\om|\lp^2\, r - \gamma |\om|\lp^2 G M = 0$, which, in the limit of large $M$, approximately becomes: $r^3 -2G M r^2 - \gamma |\om|\lp^2 G M \simeq 0$. The last term is suppressed by the factor $\lp^2$, so finally the large $M$ limit of the horizon equation is: $r^3 -2G M r^2 \simeq 0$, whose positive solution is $r=2G M$. Therefore we look for an approximate solution of Eq.\eqref{H} by perturbing around the classical Schwarzschild solution, namely we pose $r=2GM + \ve$, and we keep only the first order terms in $\ve$.}
In so doing, we get a linear equation for $\ve$, from which results
\be
\ve = \frac{(2+\gamma)|\om|}{4}\frac{\hbar}{M}\,,
\ee  
and finally the behavior of the horizon for large $M$, $M\to \infty$, is
\be
r_h \ \simeq \ 2GM \ + \ \frac{(2+\gamma)|\om|\hbar}{4M}\,,
\label{rh}
\ee 
so the usual Schwarzschild expression for the horizon is recovered in the large $M$ limit.
 
We can also investigate the behavior of the horizon in the small $M$ limit. 
%We will see that, although mathematically doable and sound, the result will be essentially meaningless from the physical point of view. 
For $M\to 0$, Eq.\eqref{H} reads 
\be
r^3 - |\om|G\hbar \,r \ \simeq \ 0  \quad \Rightarrow \quad r(r^2 - |\om|G\hbar) \ \simeq \ 0\,.
\ee
The last equation has solutions $r_{1/2}=\mp\sqrt{|\om|G\hbar}$ and $r_3=0$. We can perturb around these solutions by posing $r_{1/2}=\mp\sqrt{|\om|G\hbar}+\ve$, $r_3=0+\ve$ and setting them back into Eq.\eqref{H}. Under the hypothesis 
\be
|\ve| \ll \sqrt{|\om|G\hbar}
\label{cond}
\ee
we can retain the first order in $\ve$ only, then we arrive at a linear equation in $\ve$, and from that finally we get
\be
r_{1/2} \ \simeq \ \mp \sqrt{|\om|G\hbar} \ + \ \left(1 + \frac{\gamma}{2}\right)G M\,; \quad \quad \quad \quad
r_3 \ \simeq \ -\gamma G M\,.
\ee
Clearly, the only acceptable solution is $r_2$, since $r_2>0$ always. So we have the behavior of the horizon for small $M$ as
\be
r_h \ \simeq \ \sqrt{|\om|G \hbar} \ + \ \left(1 + \frac{\gamma}{2}\right)G M\,.
\label{rhsmall}
\ee
However from the physical point of view, this solution is practically meaningless, since the condition \eqref{cond} implies
\be
G M \ll \sqrt{|\om|G\hbar} \sim \lp\,,
\ee 
which means that this solution is valid when $M \ll m_p$, i.e. when the collapsing mass is much smaller than the Planck mass. Equivalently, the correction $\ve$ should be smaller than the Planck length, which again would mean that $\ve$ has no definite physical meaning.

%%%%%%%%%%%%%%%%%%%%%%
\subsection{Singularity $r_0$}
%%%%%%%%%%%%%%%%%%%%%%

As we have seen, the singularity of the metric \eqref{negom} is located at the only positive root $r=r_0$ of the equation 
\be
r^3 -  |\om| G \hbar r - \gamma |\om| G^2 \hbar M = 0
\label{S}
\ee
Here also we have a look at the behavior of the solution for $M$ small and for $M$ large. 

For $M\to 0$, the equation is again of the form 
\be
r^3 -  |\om| G \hbar r = 0\,,
\ee 
which has the exact solutions $r_{1/2}=\mp\sqrt{|\om|G\hbar}$ and $r_3=0$. Perturbing around them, namely posing 
\be
r_{1/2}=\mp\sqrt{|\om|G\hbar}+\ve\,, \quad \quad \quad r_3=0+\ve\,,
\ee
and substituting them back into \eqref{S}, we get, under the usual condition $|\ve| \ll \sqrt{|\om|G\hbar}$,
\be
r_{1/2} \ \simeq \ \mp \sqrt{|\om|G\hbar} \ + \frac{\gamma}{2}G M\,; \quad \quad 
r_3 \ \simeq \ -\gamma G M\,.
\ee
Repeating similar considerations as before we can say that the only acceptable solution is $r_2$ (because $r_2>0$ always), so the behavior of the radial coordinate of the singularity for small $M$ is
\be
r_0 \ \simeq \ \sqrt{|\om|G\hbar} \ + \ \frac{\gamma}{2}G M\,.
\label{r0small}
\ee
However, the above condition on $\ve$ once again implies $|\ve| \ll \lp$, or $M \ll m_p$, which makes such mathematical solutions of little, if any, physical interest. 

Instead, for $M\to \infty$ the situation is much more interesting. The large $M$ limit of the singularity equation \eqref{S} is 
\be
r^3 \ - \ \gamma |\om| G^2 \hbar M \ \simeq \ 0\,,
\ee    
so the real positive solution is $r=(\gamma |\om| G^2 \hbar M)^{1/3}$. We perturb around it by posing
\be
r \ = \ (\gamma |\om| G^2 \hbar M)^{1/3} \ + \ \ve
\ee
and under the usual condition $|\ve| \ll (\gamma |\om| G^2 \hbar M)^{1/3}$, we get for large $M$
\be
r_0 \ \simeq \ (\gamma |\om| G^2 \hbar M)^{1/3} \ + \ \frac{|\om| G \hbar}{3(\gamma |\om| G^2 \hbar M)^{1/3}}\,.
\label{r0}
\ee
An important physical consideration can now be stated. As it was already suggested by the initial analysis, the singularity results to be always protected by the horizon. Namely, by comparing Eqs.\eqref{rh}, \eqref{r0} for large 
$M$, or instead by comparing Eqs.\eqref{rhsmall}, \eqref{r0small} in the small $M$ limit, we always have 
\be
r_0 \ < \ r_h\,,
\ee
so there are no naked singularities.\\
For sake of clarity, in Fig.3 the reader can find the plots of the mass function $M(r_h)$ for the horizon (red dashed line)
\be
GM(r_h) = \left. \frac{r^3 - |\om|\lp^2 r}{2r^2 + |\om|\lp^2\gamma} \right|_{r=r_h}\,,
\label{hmf}
\ee  
the mass function $M(r_0)$ for the singularity (blue dot-dashed line)
\be
GM(r_0) = \left. \frac{r^3 - |\om|\lp^2 r}{|\om|\lp^2\gamma} \right|_{r=r_0}\,,
\ee
and the standard Schwarzschild horizon $GM=r_{SCH}/2$ (green solid line). Any horizontal line (black dashed) representing an arbitrary $M>0$ intersects first the blue line and then the red line, namely $r_0(M)<r_h(M)$ for any $M>0$. Notice that both the horizon and the singularity mass functions have a simple zero at 
\be
r = r_c = \sqrt{|\om|}\,\lp
\ee
with $r_c^2 = |\om|\,\lp^2$.

%%%%%%%%%%%%%%%%%%%%%%%%%%%%%
%
\begin{figure}[b]
	%\hspace{-2mm}
	\includegraphics[scale=.4]{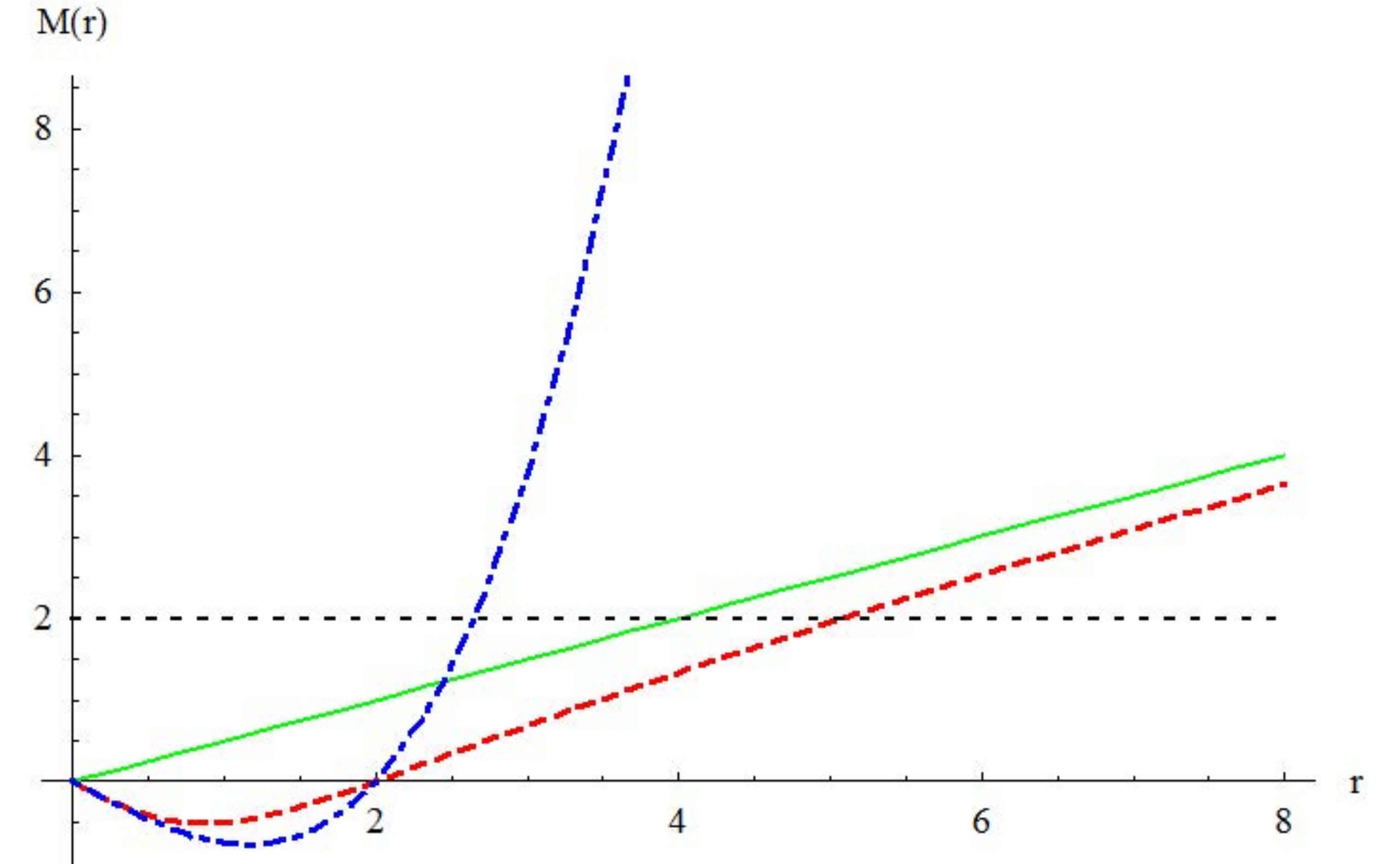}
	\centering
	\protect\caption{Horizon mass function $M(r_h)$ (red dashed line), singularity mass function $M(r_0)$ (blue dot-dashed line), and Schwarzschild mass function (green solid line). The horizontal black dashed line represents an arbitrary $M>0$, and always intersects blue and red lines at $r_0(M)<r_h(M)$, namely there are no naked singularities.}
	\label{fig3}
\end{figure}
%%%%%%%%%%%%%%%%%%%%%%%%%%%%%%%

%%%%%%%%%%%%%%%%%%%%%%%%%%%%%%%%%%%
\section{A metric for the Planck stars}
%%%%%%%%%%%%%%%%%%%%%%%%%%%%%%%%%%%

It is of the greatest interest to examine the core of the black hole metric we obtained. For masses $M$ larger than the Planck mass $m_p$, $M\gg m_p$, the central hard singularity is located at $r=r_0$, and it has, in fact, a finite positive size $r_0 \simeq (\gamma |\om| G^2 \hbar M)^{1/3} > 0$, contrary to what happens in the standard Schwarzschild black hole, where the singularity is point-like. Observe that this finite size is completely of quantum origin: in fact, $r_0 \to 0$ if we take the classical limit $\hbar \to 0$. 

But most importantly, if we presume that the whole collapsing mass $M$ is concentrated into the central hard sphere of radius $r_0$, then we can compute the (non covariant) volume of this sphere, and hence the density of this matter (as seen by an observer at infinity), which will result to be finite, and precisely
\be
r_0 = (\gamma |\om| \lp^2 G M)^{1/3} \ = \ (\gamma|\om|/2)^{\frac{1}{3}} \left(\frac{M}{m_p}\right)^{\frac{1}{3}}\lp \ \quad \ \Rightarrow \ \quad \ \varrho = \frac{M}{V_{core}} = \frac{3}{2\pi}\frac{m_p}{\gamma |\om| \lp^3} \simeq 
\frac{m_p}{2\gamma|\om| \lp^3} = \frac{\varrho_{Planck}}{2\gamma|\om|},
\label{size}
\ee
where we used the definitions $G\hbar = \lp^2$, $2G m_p = \lp$, and    
\be
\varrho_{Planck} = \frac{m_p}{\lp^3}\,.
\ee
So the central hard kernel of our black hole results to have a \textit{finite} size, and a density of the order of the Planck density. These are exactly the characteristics of the so called Planck stars, first proposed in Ref.~\cite{RovelliPS}, on the ground of general qualitative considerations. Many of the general properties described in~\cite{RovelliPS} (see also Ref.~\cite{BonannoCasadio}) can now be repeated for our black hole. The finite positive size of the central core, being of pure quantum origin, is presumably due to the action of the Heisenberg uncertainty principle, which prevents matter to be arbitrarily concentrated into a geometrical point of size zero. The central kernel can presumably keep trace of all the information swallowed by the black hole: we see here a possible way out of the information paradox. Of course, all the above considerations make sense only for $\gamma >0$ strictly.  

The original proposal~\cite{RovelliPS} contains a certain amount of qualitative considerations, including an educated guess on the form of the metric able to describe a Planck star. Such metric was initially chosen to be the Hayward metric \cite{hayward}
\be
F(r) = 1- \frac{2 G M r^2}{r^3 + 2GML^2}
\label{hay}
\ee
where $L$ is a parameter with dimensions of a length. No particularly compelling argument, from the physical point of view, was exhibited for that choice, with the exception, perhaps, that the Hayward metric is a well known example of singularity-free metric. For large $M$ the metric \eqref{hay} develops two horizons, one inner
\be 
r_{-} \ \simeq \ L \ + \ \frac{L^2}{4GM}\,,
\label{r-}
\ee 
and one outer 
\be
r_{+} \ \simeq \ 2GM \ - \ \frac{L^2}{2GM}. 
\ee
However, no specific indication is contained in the Hayward metric \eqref{hay} about the size of the hard kernel of a Planck star. Even identifying such hard kernel size with $r_-$, certainly it does not increase in size with $M$, as instead Eq.\eqref{size} suggests for $r_0$ (compare with Eq.\eqref{r-}). Even worse, the Hayward metric \textit{per se} is unable to mimic the well established quantum correction to the Newtonian potential~\cite{dono2} that occurs at low energies. This is due to the lack of a term $1/r^3$ in the expansion of the metric \eqref{hay}. 
The authors of Ref.~\cite{delorenzo} found a smart way to cure this shortcoming, but at the price of introducing a further metric function $H(r)$, determined through a bunch of additional constraints, so that their "modified Hayward" metric now reads
\be
ds^2 = -\ H(r)F(r)dt^2 + \frac{1}{F(r)}dr^2 + r^2d\Omega^2\,, 
\quad {\rm with} \quad H(r) = 1-\frac{\beta GM \alpha}{\alpha r^3 + \beta GM}\,, 
\ee
where $\beta$ is a parameter that in Ref.\cite{delorenzo} plays the r\^ole of our $|\om|$.
The above metric \footnote{A further requirement imposed by the authors on the function $H(r)$ is that $H(r)$ should allow for a time delay between a clock sitting at the center of the collapsed object ($r=0$), and a clock at infinity (put a clock at $r=0$ is in principle conceivable, just because the Hayward metric is regular at $r=0$). To get this, authors demand that 
$H(r=0)=1-\alpha$. They justify this further request by saying that "is a physically unmotivated restriction" to leave $H(0) = 1$. 
%{\cdred We disagree with this point of view. In fact, an easily conceivable model is for example a spherical shell of (null) matter, empty inside: the metric then will be Minkowski inside, and Schwarzschild outside, and there is no time delay between a clock at the center of the shell and one at infinity.} 
In any case, we do not have this kind of problem with the SDG metric \eqref{negom}, since the center $r=0$ cannot be reached, being protected by the singularity at $r=r_0>0$.} finally contains the $1/r^3$ term necessary to mimic the Donoghue modified Newtonian potential~\cite{dono2} for large $r$. \\
Although smart and working, the above solution is undeniably contorted and intricate. On the contrary, within the formalism of the 'renormalization improved' and 'running' coupling constants, the mathematical structure of the metric is dictated by the general properties of the SD/AS Gravity, and its lapse function \eqref{negom} results clearly simpler than the above product $H(r)F(r)$. Our SDG metric \eqref{negom} already contains the right terms to match, at large distances, the quantum corrected Newtonian potential. Moreover, and this is quite astonishing, by simply imposing that match, the final form of the metric is uniquely fixed, and it automatically displays the correct size of the central hard kernel of the Planck star.

In the following, we shall therefore study the thermodynamic properties of the metric \eqref{negom}.

%[coincide con la previsione di Rovelli m/m_P^1/3]

%[from an educated guess to a firm prediction of the ASG formalism] 

%[1. noi abbiamo una predizione esatta del size della planck star e voi no;\\ 
%2. noi abbiamo una metrica esatta e voi no] 

%%%%%%%%%%%%%%%%%%%%%%%%%%%
\section{Hawking temperature}
%%%%%%%%%%%%%%%%%%%%%%%%%%%

According to the celebrated works of Hawking (see e.g. Ref.~\cite{Hawking}), a generic Lorentzian black hole metric as \eqref{metric} with a lapse function $F(r)$ bearing an horizon (namely a simple zero at some $r=r_h$, with $F(r_h)=0, F'(r_h)\neq 0$), displays on the horizon a Hawking-Bekenstein temperature given by
\be
T_{BH}=\frac{\hbar}{4\pi}F'(r_h)\,.
\ee 
Therefore, reminding $G\hbar=\lp^2$ and $|\om|\lp^2 = r_c^2$, from \eqref{negom} we can compute,  
\be
\frac{4\pi}{\hbar} T \ = \ F'(r) \ = \ 2GM r \,\frac{r^3 + r_c^2 r + 2\gamma GMr_c^2}
{(r^3 -  r_c^2 r - \gamma GMr_c^2)^2}
\ee
where we take $r=r_h$, namely $r$ should satisfy the identity $F(r)=r^3 - 2GM r^2 - r_c^2 r - \gamma GMr_c^2=0$.\\ 
We can use this to semplify $r^3 - r_c^2 r - \gamma GMr_c^2 = 2GMr^2$, and get
\be
F'(r) \ = \ \frac{1}{2GM} \left(1 \ + \ \frac{r_c^2}{r^2} \ + \ \frac{2\gamma GM r_c^2}{r^3}\right)\,.
\label{F'}
\ee  
The above expression is exact. Now we study its limits, namely the limits of $T(M)$ for $M$ large and for $M$ small. \\

Repeating twice the procedure which led us to \eqref{rh}, for $M \to \infty$, we can arrive to write the expansion of $r_h(M)$ to the third order in $1/M$ 
\be
r \ = \ r_h \ \simeq \ 2GM \ + \ \frac{(\gamma +2)r_c^2}{4GM} \ + \ \frac{(\gamma +2)r_c^4}{16 G^3 M^3} \ + \ \dots
\label{rexp}
\ee
Inserting this back into \eqref{F'} we get finally
\be
\frac{4\pi}{\hbar} T(M) \ = \ F'(r) \ = \ \frac{1}{2GM}\left[1 + \frac{\gamma+1}{4}\left(\frac{r_c}{GM}\right)^2 - \frac{(\gamma+2)(3\gamma+2)}{32}\left(\frac{r_c}{GM}\right)^4 + \dots\right]\,.
\label{TMlarge}
\ee
We recover here the standard behavior $T_{BH}=\hbar/8\pi GM$ of the BH temperature for large $M$, when our black hole looks even more like a Schwarzschild one. Moreover, notice that the horizon mass function \eqref{hmf} is an odd function $M(r_h)$, therefore its inverse $r_h(M)$ is an odd function, and its expansion can hence contain odd powers of $M$ only, as in fact results in Eq.\eqref{rexp}. Besides, the expression in round brackets in \eqref{F'} must be an even function of $M$, since $r_h(M)$ is odd, and therefore we find in its expansion \eqref{TMlarge}, square brackets, only even powers of $1/M$. \\ 

For small $M$, first we note that, clearly, the horizon mass function $M(r_h)$, diagram Fig.3, does not have a positive minimum, $M_{min}>0$. So the Hawking evaporation can in principle proceed until $M \to 0$. Reminding Eq.\eqref{rhsmall}, we can write for small $M$
\be
r_h(M) \ = \ r_c \ + \ \left(1+\frac{\gamma}{2}\right)GM \ + \ \dots
\ee 
where we suppose $GM\ll r_c$. Inserting this $r_h(M)$ back into \eqref{F'} we get
\be
\frac{4\pi}{\hbar} T(M) \ = \ F'(r) \ = \ \frac{1}{2GM}\left[2 + (\gamma-2)\left(\frac{GM}{r_c}\right) + \frac{3(\gamma+2)(2-3\gamma)}{4}\left(\frac{GM}{r_c}\right)^2 + \dots\right]\,.
\label{TMsmall}
\ee
We see here again a behavior similar to that of standard Hawking temperature for the standard Schwarzschild black hole, namely $T(M) \to \infty$ when $M \to 0$. However, for small $M$ we see that our SDG-modified black hole displays a temperature double of that of the standard Hawking, namely $T(M) \simeq 2 \, T_{BH}(M)$ for small $M$. \\

At this point, for the forthcoming developments, it is useful to observe that for a metric with a lapse function $F(r,M)$, the Hawking temperature can be expressed in two different ways, either $T$ as a function of $M$, $T=T(M)$, or instead $T$ as a function of $r$, $T=T(r)$. In fact, usually we write
\be
\frac{4\pi}{\hbar} T(M) \ = \ \left.\frac{\partial F(r,M)}{\partial r}\right|_{r=r(M)}\,,
\ee   
where $r(M) \equiv r_h$ is a solution of the equation $F(r(M),M)=0$. However, the equation $F(r,M)=0$ in respect to $r$ can be complicated, containing radicals, etc., as actually it is in our present case, where we have the third degree equation~\eqref{H}. It is much easier to consider $T$ as a function of $r$, as
\be
\frac{4\pi}{\hbar} T(r) \ = \ \left.\frac{\partial F(r,M)}{\partial r}\right|_{M=M(r)}\,,
\ee
where $M(r)$ is a solution of $F(r, M(r))=0$, namely $M(r)$ is the mass function, $M(r) \equiv M(r_h)$. Mathematically, $r=r(M)$ and $M=M(r)$ are just the same implicit function, locally defined by the equation $F(r,M)=0$. Of course, the equation $F(r,M)=0$ is usually much easier to be solved in respect to $M$ than in $r$, being usually an equation of first degree in $M$.  

Following the above considerations, we insert the mass function espression \eqref{hmf} into Eq.\eqref{F'}, and we get
\be
\frac{8\pi}{\hbar} T(r) \ = \ \left(\frac{2r^2 + \gamma\, r_c^2}{r^3 - r_c^2\, r}\right)\left(1+\frac{r_c^2}{r^2}\right) \ + \ \frac{2\gamma\, r_c^2}{r^3}\,,
\label{Tr}
\ee
where as usual $r_c^2=|\om|\lp^2$, and $r \equiv r_h$ is the only real positive solution of the horizon equation \eqref{H}. As expected, we obtain here an exact, simple, rational expression of $T$ as function of $r$, with no radicals displayed. 
Notice that we can compute the behavior of $T(r)$ for large $r$ (corresponding to $M\to\infty$), or for $r\to r_c$ (corresponding to $M\to 0$), and we get respectively
\be
T(r) \ \simeq \ \frac{\hbar}{4\,\pi\, r} \quad \ {\rm for} \ \quad r\to \infty; 
\quad \quad \quad T(r) \to \infty \quad \ {\rm for} \ \quad r\to r_c\,.
\ee
The above confirms the behavior of $T(M)$ showed, respectively, in Eqs. \eqref{TMlarge}, \eqref{TMsmall}.

%%%%%%%%%%%%%%%%%%%%%%%%%%%
\section{Specific heat}
%%%%%%%%%%%%%%%%%%%%%%%%%%%

We can now proceed swiftly to the computation of the specific heat capacity. In general, it is defined as $C_s=dE/dT$, where $E$ is the total energy of the system under consideration. As usual we identify the total energy of our SDG black hole with its total mass $M$ ($c=1$) (see Ref.\cite{Bonanno:2000ep}). Once again, as above, the 'best' analytical parameter through which express the mass and the temperature of our black hole is not its mass $M$, but rather its gravitational radius $r_h=r$ (positive solution of Eq.\eqref{H}). So we have 
\be
C_s = \frac{dM(r)}{dT(r)} = \frac{M'(r)}{T'(r)}\,.
\ee 
For agility of calculation, let's rename the horizon mass function \eqref{hmf} as
\be
GM(r) \ = \ \frac{r^3 - r_c^2 r}{2r^2 + \gamma r_c^2} \ =: \ N(r)\,,
\label{N(r)}
\ee
with $r_c^2=|\om|\lp^2$, then Eq.\eqref{Tr} reads
\be
\frac{8\pi}{\hbar} T(r) \ = \ \frac{1}{N(r)}\left(1+\frac{r_c^2}{r^2}\right) \ + \ \frac{2\gamma\, r_c^2}{r^3}\,.
\label{TNr}
\ee
Therefore we have
\be
C_s \ = \ - \frac{8\pi}{G\hbar} \ \frac{N'(r)N(r)^2}{\left[N'(r)(1+r_c^2/r^2) + 2N(r)r_c^2/r^3 + 6N(r)^2\gamma r_c^2/r^4\right]}\,.
\ee
We can easily verify that $N(r) > 0$, and $N'(r) > 0$, for $r > r_c$ (being $\gamma \geq 0$). Therefore
\be
C_s(r) < 0 \quad \quad {\rm for \ any} \quad \quad r>r_c \,,
\ee
namely the specific heat is negative for any $r>r_c$, i.e. when $M>0$. This behavior is analogous to the Schwarzschild black hole. More specifically, since $N(r_c)=0$ and $N'(r_c)=2/(\gamma+2)>0$, we have $C_s(r_c)=0$, or, more precisely,
\be
C_s(r) \ \simeq \ - \frac{4\pi}{G\hbar}\left(\frac{2}{2+\gamma}\right)^2 (r-r_c)^2 \ + \ \dots\,.
\ee
Summarizing, when $r \to r_c$ then
\be
M(r) \to 0, \quad \quad T(r) \to \infty, \quad \quad C_s(r) \to 0 \,,
\ee 
in perfect analogy with the Schwarzschild black hole.\\
When instead $r \to \infty$, we have $r \simeq 2GM$ and therefore
\be
C_s(r) \ \simeq \ - \frac{2\pi}{G\hbar}\, r^2 \ \simeq \ - \frac{8\pi}{\hbar}\, GM^2
\ee
which coincides with the behavior of $C_s$ for large $M$ for the Schwarzschild black hole.

%%%%%%%%%%%%%%%%%%%%%%%%%%%
\section{Emission rate equation}
%%%%%%%%%%%%%%%%%%%%%%%%%%%

We can investigate how long it takes a black hole with an initial mass $M$ to reduce to a final mass $M_f$ via the Hawking radiation. In our case we have seen that the black hole can evaporate until $M_f=0$. The Stefan-Boltzmann law allows us to write an emission rate differential equation, which, once integrated, yields the above life-time for the black hole considered. The mass/energy loss per unit proper time of an infinitely far away static observer is approximately given by 
\be
-\frac{dM}{dt} \ = \ \sigma \,{\cal A} \,T^4\,,
\ee   
where $\sigma$ is a constant (related to the Stefan-Boltzmann constant) and $\cal A$ is the area of the event horizon. In the standard Hawking calculation for a Schwarzschild black hole we have ${\cal A}\sim r_h^2 \sim M^2$, and $T\sim 1/M$. Therefore we get a life-time $t_{life} \sim M^3$. As before, for the description of our system the variable $r \equiv r_h$ (positive solution of the horizon equation \eqref{H}) appears analytically more viable than the mass $M$. So we consider everything as a function of $r$, namely $M(r)$, ${\cal A}(r)$, $T(r)$, and the above equation becomes an evolution equation for the gravitational radius
\be
-\frac{dr}{dt} \ = \ \sigma \,\frac{{\cal A}(r)\,T(r)^4}{M'(r)}\,.
\ee 
Inserting ${\cal A}(r)=4\pi r^2$, and $M(r)$, $T(r)$ from Eqs.\eqref{N(r)}, \eqref{TNr}, respectively, we arrive at 
\be
-\frac{dr}{dt} \ = \ 4\pi G\sigma \left(\frac{\hbar}{8\pi}\right)^4 \frac{r^2}{N'(r)}
\left[\frac{1}{N(r)}\left(1+\frac{r_c^2}{r^2}\right) \ + \ \frac{2\gamma\, r_c^2}{r^3}\right]^4 \,.
\label{ERE}
\ee 
This equation doesn't look very expressive, although it can be integrated, in principle, without a particular effort (a rational function, quite tedious calculation). However, it can be used to check the two important limits of our physical system, namely $r \to \infty$ and $r \to r_c$. \\
When $r \to \infty$, then $N(r) \simeq r/2$ and $N'(r) \simeq 1/2$. Hence
\be
-\frac{dr}{dt} \ \simeq \ \frac{\sigma G \hbar^4}{32\pi^3} \ \frac{1}{r^2} \quad \quad {\rm for} \quad \quad r \to \infty\,.
\ee
Since in this approximation $r \sim M$,  then we recover here the standard Hawking result, namely $-dM/dt \sim 1/M^2$. Therefore, in this approximation, for the life-time of black hole of initial mass $M$, the above equation once integrated gives $t_{life} \sim M^3$, as expected. \\
When $r \to r_c$, from Eq.\eqref{N(r)} we get the behavior of $N(r)$ as 
\be
GM(r)=N(r) \ \simeq \ \frac{2}{2+\gamma}(r-r_c) + \dots 
\label{Nsr}
\ee
%we know that $N(r) \to 0$, and $N'(r) \to 2/(2+\gamma)$. 
Therefore the RHS of the emission rate equation \eqref{ERE} diverges badly as, 
\be
-\frac{dr}{dt} \ \simeq \ 2\pi G \sigma \left(\frac{\hbar}{8\pi}\right)^4 \ \frac{(2+\gamma)^5 r_c^2}{(r-r_c)^4} \ \to \ +\infty \,,
\ee
which can be expressed in terms of the $M \to 0$ as
\be
-\frac{dM}{dt} \ \simeq \ \frac{\sigma\hbar r_c^2}{G}\left(\frac{\hbar}{4\pi G}\right)^3 \frac{1}{M^4}\,.
\label{1/M4}
\ee
Here the divergence is worse than in the standard Hawking calculation, where $-dM/dt \sim 1/M^2 \to +\infty$ when $M \to 0$.  Although trusting Eq.\eqref{1/M4} to its very end appears risky and perhaps incorrect, nevertheless its behavior signals as well an explosive character of the last instants of life of a SDG-modified black hole.

%%%%%%%%%%%%%%%%%%%%%%%%%%%
\section{Entropy of the Planck star} %quantum black hole}
%%%%%%%%%%%%%%%%%%%%%%%%%%%

We consider here the computation of the thermodynamic entropy of our system, while we leave to future work any possible statistical mechanics interpretation of such an entropy through the counting of "microscopic" states (maybe inaccessible to our observation). Therefore, by identifying, as usual, the energy of our system with the mass $M$ of the black hole ($c=1$), we can write from general thermodynamics
\be
dS \ = \ \frac{dM}{T}\,,
\ee  
where of course $T$ is the Hawking temperature of the hole.\\
The standard calculation proceeds by considering $T$ as a function of $M$, and then computing $T(M)$, $S(M)$. On the contrary in our case, as widely illustrated above, the positive solution $r$ of Eq.\eqref{H}, namely the radius of the event horizon, is an analytical variable better viable than $M$. Therefore we shall write
\be
dS \ = \ \frac{M'(r)}{T(r)}\,dr\,.
\ee
Hence, using Eqs.\eqref{N(r)}, \eqref{TNr}, we can write
\be
\frac{dS}{dr} \ = \ \frac{M'(r)}{T(r)} \ = \ \frac{8\pi}{G\hbar} N'(r)N(r)\left[1+\frac{r_c^2}{r^2} + \frac{2\gamma r_c^2}{r^3}N(r)\right]^{-1} \,. 
\label{Sr}
\ee
In principle, as before, we can integrate exacly the above equation, but this step doesn't seem to give us any particularly useful information. It appears more clever to expand the above expression around the two significant limits $r\to\infty$ and $r\to r_c$, and then integrate.\\
For $r \to \infty$,  then $N(r) \simeq r/2$ and $N'(r) \simeq 1/2$. Hence
\be
\frac{dS}{dr} \ = \ \frac{2\pi}{G\hbar} \ r\,,
\label{Slr}
\ee
which can be integrated to give
\be
S(r) - S(\Lambda) = \int_\Lambda^r \frac{2\pi}{G\hbar} \ \rho d\rho = \frac{4\pi r^2}{4\lp^2} - \frac{4\pi \Lambda^2}{4\lp^2}\,,
\ee
where we inserted a large (infrared) cutoff $\Lambda$ to take into account the fact that we are integrating Eq.\eqref{Sr} in a region of large $r$ (where \eqref{Sr} takes the form \eqref{Slr}). Once again, we recover for large $r$ the well known behavior of the entropy of a Schwarzschild black hole
\be
S(r) \ \sim \ \frac{{\rm Area}}{4\lp^2}\,.
\ee 
When on the contrary $r\to r_c$, then Eq.\eqref{Nsr} holds, and $N'(r)\simeq 2/(2+\gamma)$, hence
\be
\frac{dS}{dr} \ \simeq \ \frac{4\pi}{G\hbar}\left(\frac{2}{2+\gamma}\right)^2 (r-r_c)
\ee 
which integrated yields
\be
S(r) - S(r_c) \ \simeq \ \frac{2\pi}{G\hbar}\left(\frac{2}{2+\gamma}\right)^2 (r-r_c)^2\,.
\ee
$S(r_c)$ could be interpreted as the entropy of the central core (after complete evaporation), which may in the end account for the information swallowed by the black hole, and therefore perhaps represent a way out of the information paradox (although a 'zero mass' remnant remains at the moment a puzzling object).   
\section{Conclusions}
%%%%%%%%%%%%%%%%%%%%%%%%%%

%\textbf{\color{blue} DA RISCRIVERE MEGLIO - 
In this paper we have described the main features of a spherically symmetric black hole metric suggested by a Scale Dependent Gravity (SDG) approach. Respect to previous studies carried out in the Asymptotically Safe Gravity (ASG) framework, the decisive novelty of the present work is that we investigated a \textit{negative} value of the free parameter $\om$. This was suggested by a comparison between  the SDG/ASG corrected Newtonian potential, with the analog quantum corrections recently computed by Donoghue, Khriplovich, and collaborators~\cite{dono2,kiril2,dono3} using the approach to GR as a low energy effective QFT.  

The fact $\om < 0$ completely changes the geometry of the SDG 'modified' black hole metric. Previously unexplored aspects of this metric have been studied, the most relevant one being the presence of a finite-size singularity at the core of the black hole. Surprisingly, the size of this "black core" results to be exactly what needed to describe the so called Planck stars. These objects were introduced years ago on the basis of semi-qualitative arguments~\cite{RovelliPS}, while in our context they appear as a quite natural mathematical consequence of the SDG metric with a negative $\om$ parameter (see also e.g.~\cite{HHK}).

Hawking temperature, specific heat, emission rate equation, and thermodynamic entropy have been studied for our Planck star metric, and they yield illuminating insights. It is worth mentioning that the phenomenology of these objects could be quite rich, and presents both astrophysical and cosmological signatures, in particular in the realm of (primordial) black hole evaporation~\cite{AAVV}. Since a Planck star evaporates with a hard kernel of finite positive size, then the final explosion may occur at a "macroscopic" scale, namely at a much bigger scale than the Planck scale. So, Planck star explosions could be naturally associated with some of the measured short gamma-ray bursts (SGRBs)~\cite{nakar}. In Ref.~\cite{Barrau-Rovelli} authors estimated that several short gamma-ray bursts per day, around 10 MeV, with an isotropic distribution, can be expected coming from a region of a few hundred light years around us. On the other hand, also fast radio bursts, strong signals with millisecond duration, which are probably of extragalactic origin, have been shown in Ref.~\cite{BRV} to have wavelengths not far from the expected size of the exploding hole. 

%{\cdred For example, burst in the CMB...}

On the theoretical side, further investigations related to this kind of metrics are currently being carried out, aimed to better understand Penrose diagrams, energy conditions, singularity theorems, quasi-normal modes, as well as a statistical interpretation of entropy and information paradox.

\section*{Acknowledgements} 

The authors sincerely thank A.~Bonanno, M.~Reuter, A.~Platania, R.~Casadio, A.~Flachi, M.~Fr\"ob and an anonymous referee for helpful suggestions and enlightening comments.
%%%%%%%%%%%%%%%%%%%%%%%%%%%

%%%%%%%%%%%%
%
%
%
\end{document}